# Electronic coupling of colloidal CdSe nanocrystals monitored by thin-film positron-electron momentum density methods

S.W.H. Eijt,[1,a)] P.E. Mijnarends,[1,4] L.C. van Schaarenburg,[1] A.J. Houtepen,[2] D. Vanmaekelbergh,[3] B. Barbiellini,[4] and A. Bansil[4]

[1]*Department of Radiation, Radionuclides and Reactors, Faculty of Applied Sciences, Delft University of Technology, Mekelweg 15, NL-2629 JB, Delft, The Netherlands*

[2]*DelftChemTech, Faculty of Applied Sciences, Delft University of Technology, Julianalaan 136, NL-2628 BL, Delft, The Netherlands*

[3]*Institute of Physics and Chemistry of Nanomaterials and Interfaces, Utrecht University, P.O. Box 80000, NL-3508 TA, Utrecht, The Netherlands*

[4]*Physics Department, Northeastern University, Boston, Massachusetts 02115*

**Abstract**

The effect of temperature controlled annealing on the confined valence electron states in CdSe nanocrystal arrays, deposited as thin films, was studied using two-dimensional angular correlation of annihilation radiation (2D-ACAR). A reduction in the intensity by ~35% was observed in a feature of the positron annihilation spectrum upon removal of the pyridine capping molecules above 200 °C in a vacuum. This reduction is explained by an increased electronic interaction of the valence orbitals of neighboring nanocrystals, induced by the formation of inorganic interfaces. Partial evaporation of the nanoporous CdSe layer and additional sintering into a polycrystalline thin film was observed at a relatively low temperature of ~486 °C.

___________________________

a) Electronic mail: s.w.h.eijt@tudelft.nl

The size and shape of colloidal II-VI semiconductor nanocrystals can be well controlled, leading to a pronounced tunability and variation in their optical and opto-electronic properties.[1-3] Their promise for applications in light-emitting diodes, solar cells and other opto-electronic devices was demonstrated in several studies.[4-6] For example, recently ultra-thin solar cells consisting of sintered nanorods (a dual set of CdSe and CdTe) were developed.[7] The (opto-)electronic properties can be further modified by structural tailoring of nanocrystal superlattices.[8-10] The electronic interaction between neighboring nanocrystals is a fundamentally important factor for composite nanocrystal devices, which determines both the transport of electron and hole charge carriers and the electronic structure of the active layers. The coupling can, in principle, be tailored by ligand manipulation or by inorganic tunneling barriers between neighboring nanocrystals.[1,11,12] Thus, innovative heterostructures are created.[12] For example, enhanced conductivity has been achieved by removal of pyridine ligands by gentle heating (150-175 °C) in a vacuum at moderate temperatures.[1,8,10] This reduces the average distance between neighboring nanocrystals to less than 2 Å, and leads to changes in the optical properties due to strong coupling.[1,8,10,11] A recent in-situ electron microscopy study on monolayers of PbSe nanocrystals indicates that this is accompanied by rotations of the nanocrystals and the formation of an inorganic interface between neighboring nanocrystals at slightly higher temperatures.[13]

Recent studies on semiconductor nanocrystals[14-17] show the potential of positron methods to study the electronic structure of nanocrystal solids, since the positron can be used as a sensitive probe for detecting the surface composition of the nanocrystals and the confinement of the upper valence electron states. In the present study, we apply high-resolution depth-sensitive positron methods[15,17-19] to show that the electronic interaction between CdSe nanocrystals, deposited as thin layers, can be monitored through observation of the electron momentum distribution of the valence states. Further, the depth-resolved positron studies provide insights into further sintering of the nanocrystal layers at higher temperatures.

Pyridine capped colloidal CdSe nanocrystals were prepared by standard solution phase synthesis[20] at 300 °C using a TOPO/HDA mixture as a solvent; the TOPO to pyridine ligand exchange was achieved by subsequent boiling in pyridine. Thin films of nanocrystals were obtained by spin-coating on borosilicate glass substrates covered by an electro-deposited 200 nm Au film. The temperature dependence of positron 2D-ACAR distributions was monitored at a 1.1 keV positron implantation energy using the POSH-ACAR facility.[15,17] The temperature was varied in-situ using a W-Al$_2$O$_3$ resistive heating plate as a sample mount in a vacuum with a pressure of ~10$^{-8}$ mbar. A momentum window of $|p| < 0.44$ a.u. was used to extract the positron $S$-parameter (see inset of Fig.1a),[15,18] which is a measure of annihilation with valence electrons, providing sensitivity to the electronic structure and the presence of open volume defects such as nanopores and vacancies. The positron Doppler broadening of annihilation radiation (511 keV) was measured using positrons with a kinetic energy in the range of 0-25 keV.[18,19] The Doppler S-parameters were normalized to the S-parameter of bulk CdSe. Optical Absorption Spectra (OAS) were collected in the range 1.0 to 6.0 eV.

Figure 1a presents the temperature dependence of the S-parameter normalized to the S-parameter of bulk CdSe in the temperature range up to 300 °C, extracted from 2D-ACAR distributions measured on a ~48 nm thin film consisting of nanocrystals with a diameter of ~3 nm. The linear increase in $S(T)$ is a consequence of thermal expansion of the CdSe lattice and the resulting temperature dependence of the electronic structure.[21,22] The S-parameter shows a stepwise increase at $T_{set}$=250 °C, induced by the detachment and removal of pyridine capping molecules from the CdSe nanocrystals. Preliminary electron microscopy results on a monolayer of CdSe nanocrystals indicate that this is accompanied by the formation of an inorganic interface between neighboring nanocrystals, which attain the same crystal orientation in domains of ~50 nm. The full process of pyridine removal, including a possibly slow out-diffusion and subsequent evaporation at the outer surface of the film, together with this initial stage of sintering occurs on a time scale of a few days, i.e., much slower than in the case of a monolayer of nanocrystals.[13] It leads to a clear and systematic increase in the S-parameter of $\frac{\Delta S}{S_0} = +1.1\%$, which remains after cooling the CdSe nanocrystal sample back to room temperature. Clearly, alignment of nanocrystals through rotations of the nanocrystals and subsequent formation of inorganic interfaces is more complex here than for the (two-dimensional) monolayers.

In Fig. 1b, the corresponding 1D-ACAR momentum distributions collected at room temperature before and after removal of the pyridine are presented in the form of ratio curves relative to bulk crystalline CdSe. Clearly, the pyridine removal and the first sintering step lead to a significant reduction (Δ) of the order of ~35% in the intensity of the peak near 1 a.u., the presence of which is characteristic for the confinement of the Se(4p) valence electron states.[14,15,23] This shows in a direct manner that the confinement of the valence electrons reduces upon removal of the pyridine ligands and the formation of an inorganic interface between nanocrystals.[23] Consequently, an electronic coupling is established between neighboring nanocrystals. Further, we observed a clear increase in the momentum density in the range 1.5 – 2.5 a.u.. This reflects the larger contribution of Cd(4d) electrons compared to the case of isolated pyridine-capped nanocrystals,[15] and is caused by positron trapping at vacancies in the newly formed, imperfect CdSe interface between neighboring quantum dots. Cd vacancies are effective positron trapping sites in CdSe, with a higher momentum density in this momentum range relative to the case of positron trapping at surfaces of CdSe nanocrystals.[24]

Optical absorption spectra, on the other hand, showed that the removal of pyridine from a similarly prepared sample heated in the same temperature range leads to a large red-shift in the direction of the band edge absorption for CdSe single crystals (Fig. 2). Optical spectroscopy therefore indicates a nearly complete disappearance of the exciton confinement, whereas the positron measurements show that the valence electrons still experience quantum confinement, albeit reduced by ~35%. This shows that the degree of electron and hole confinement in a nanocrystal array can be very different. Similar results were obtained in scanning tunneling spectroscopy studies of PbSe nanocrystal arrays.[25] Recent ab-initio simulations[12] on neighboring CdSe nanocrystals separated by a very thin CdS intermediate barrier layer are also consistent with different confinement behavior of electrons and holes. In the present case, the observed changes in the electron momentum distribution near 1 a.u. produced by pyridine removal and interface formation can be explained by a delocalization of the Se(4p) orbital and a narrowing of the band gap.[15]

Figure 3a shows that if the nanoporous ligand-free CdSe nanocrystal layer resulting from the first heating run (Fig. 1a) is heated again (this time in the range between room temperature and 580 °C), the S-parameter increases with temperature at about the same pace as in the first heating run. The stepwise increase in the range 200-250 °C related to the removal of pyridine is now absent, as expected. At a higher temperature of ~486 °C, however, a drastic decrease of about 2% in the S-parameter takes place with a time constant $\tau$ of ~10 h. This indicates a broadening of the momentum distribution. It is, in part, the result of further sintering of the 3-nm-diameter CdSe nanocrystals, as is expected to occur because of their drastically reduced melting temperature, estimated to be in the range of ~600-700 K.[2,26] The vapor pressure of the CdSe nanocrystals will consequently become high. The resulting second sintering step leads to a removal of the nanopores from the CdSe layer.

Positron depth-profiling Doppler experiments (Fig. 3b) provide further evidence for pyridine removal and sintering of particles. VEPFIT analysis[27] of the Doppler depth profiles obtained before and after the first heating run showed that the thickness of the CdSe film reduces from an initial 48 nm (estimated mass density $\rho = 3.2 \ g \cdot cm^{-3}$) to 39 nm ($\rho = 3.3 \ g \cdot cm^{-3}$), as expected because of the removal of the pyridine. The layer thickness is further reduced to ~18 nm after the final heating run (assuming $\rho = 5.6 \ g \cdot cm^{-3}$, typical for a dense CdSe layer). This shows that a significant fraction of ~22% of the CdSe has evaporated. Further, Fig. 3b shows that the S-parameter of the top layer is close to the reference value for bulk CdSe after the final heating run, indicating that a dense polycrystalline film is formed. Our measurement shows that, for parts of the sample, the CdSe nanocrystal film is actually completely removed by evaporation, exposing the underlying Au layer. Complementary X-ray reflectometry measurements, namely, clearly revealed a signature of the critical angle characteristic for the Au-air interface.

In summary, our study shows that depth-resolved positron-electron momentum density probes are capable of providing important insights into the electronic coupling of nanocrystals embedded in active layers of future generations of solar cells and (nano)-electronic devices. The contact established between the neighboring nanocrystals leads to a delocalization of the valence orbitals (important for hole transport) and a band gap narrowing. Moreover, the electronic structure becomes accessible in a manner complementary to optical and X-ray absorption spectroscopy,[11,28] which are affected by excitonic effects. Further, our study shows that in-situ depth-resolved positron annihilation investigations of nanocrystal layers can provide insights into the basic mechanisms of sintering. Depth-profiling positron-electron momentum density methods clearly show great promise for uncovering the basic mechanisms of charge transport and electronic properties in nanocrystal composite layers, superlattices and heterostructures, which are the basic building blocks and active layers for future generations of solar cells and nanoelectronic devices.

We thank J. de Roode for facilitating the in-situ heating studies, H. Schut for the Doppler broadening measurements, and A.A. van Well for advice in the X-ray reflectometry study. The work at Northeastern University was supported by the US Department of Energy, Office of Science, Basic Energy Sciences contract DE-FG02-07ER46352.

[23]     The confinement of the positron, trapped in a surface state, also contributes to the broadening of the electron-positron momentum distribution. However, our previous studies provide evidence that this effect is much smaller than the effect of electron confinement inside the 3 nm CdSe nanocrystals (Refs. 15, 17). Surfaces of the nanocrystals remain abundantly available after this first sintering step, acting as strong trapping sites. Positron de-confinement (or delocalization) therefore cannot explain the observed strong reduction of the confinement feature in the electron-positron momentum distribution.

**Figure Captions**

**Fig. 1**:

a) Temperature dependence of the positron S-parameter derived from 1D-ACAR distributions upon heating a layer of pyridine-capped CdSe nanocrystals in vacuum.

b) Room temperature 1D-ACAR momentum distributions before (solid line) and after (dashed line) removal of pyridine ligand molecules, presented as ratio curves with respect to the bulk CdSe directionally averaged 1D-ACAR distribution. The dotted line represents the estimated ratio curve for a CdSe surface.

**Fig. 2:**

(Color online) Optical absorption spectra of (i) as-deposited layers of pyridine-capped CdSe nanocrystals with sizes 3, 4.4, and 5 nm on glass substrates (thin solid, dashed and chain lines; blue); (ii) a layer of 3 nm CdSe nanocrystals after pyridine removal by heating in a vacuum in the temperature range below 300 °C (heavy solid line, red); and (iii) bulk single crystalline CdSe (heavy dashed line, black).

**Fig. 3**:

(Color online)

a) Temperature dependence of the S-parameter derived from 1D-ACAR distributions during a second heat treatment in vacuum.

b) Room temperature Doppler broadening S-parameter depth-profiles for the layer with 3 nm pyridine-capped CdSe nanocrystals deposited on a thin Au film covering the glass substrate. (i) as-deposited (closed circles, blue) (ii) after pyridine removal (open circles, red); (iii) after the second heat treatment up to 580 °C (filled squares, black).

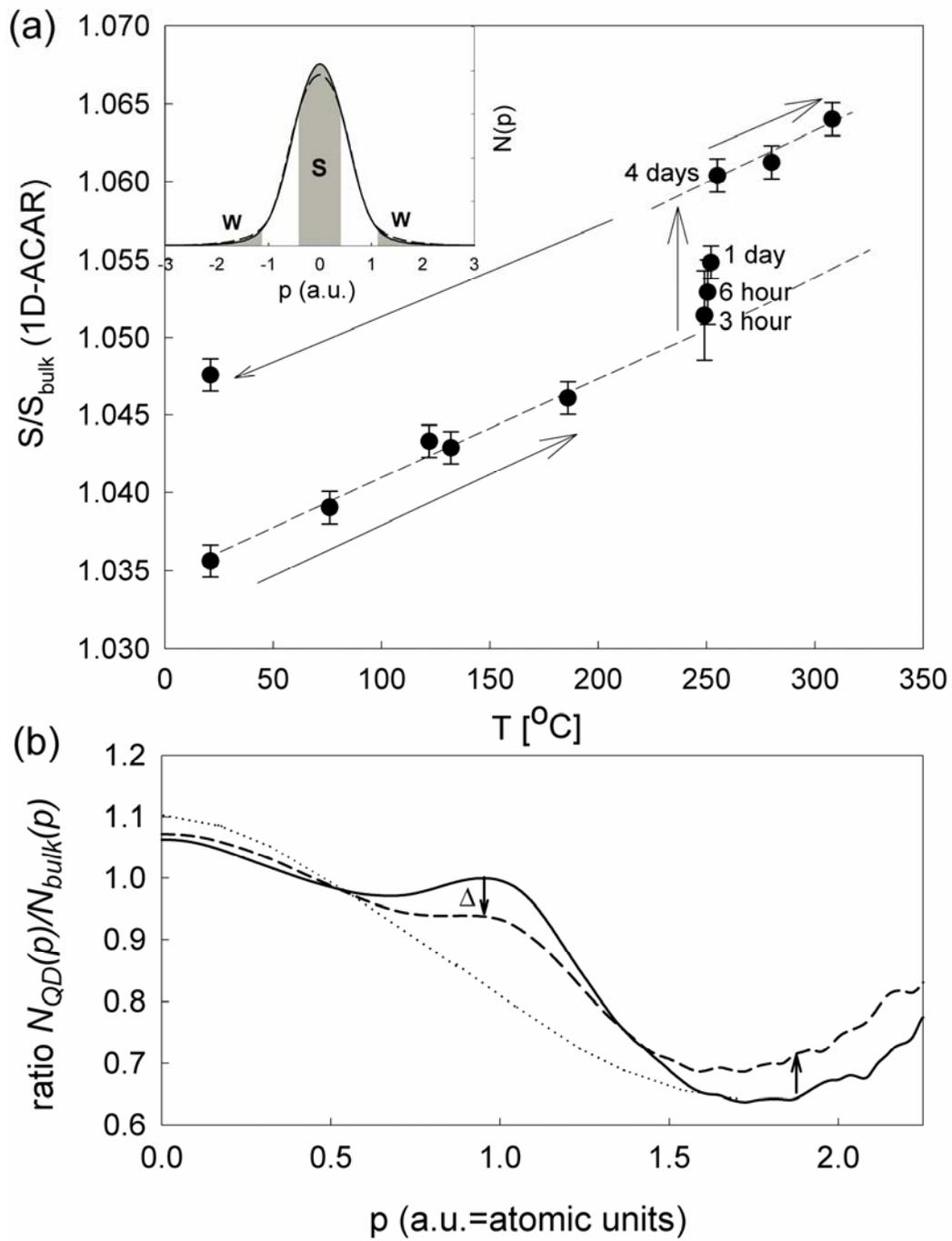

**Fig. 1**

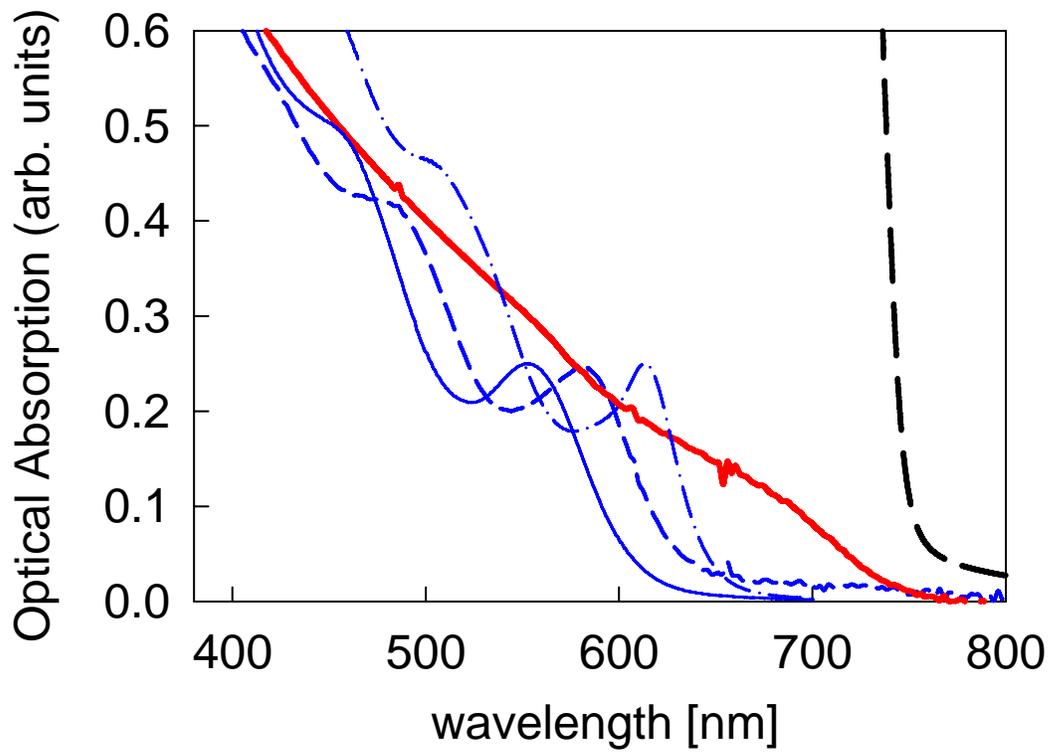

**Fig. 2**

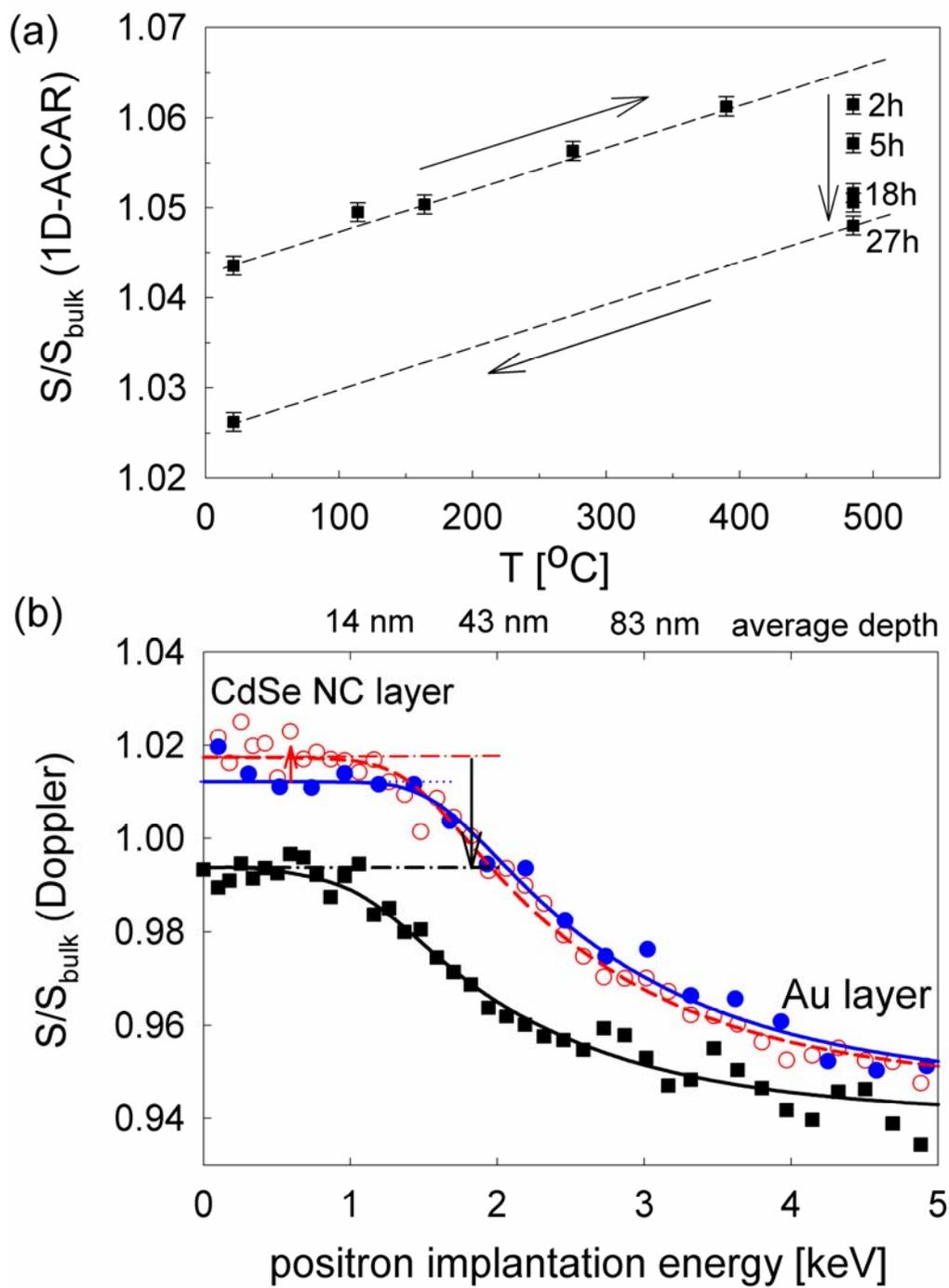

**Fig. 3**